\pdfoutput=1
\documentclass[sigconf]{acmart}

\usepackage[english]{babel}

\usepackage{amsfonts}
\usepackage{bbm}
\usepackage{bm}
\usepackage{mathtools}
\usepackage{scalerel} %

\usepackage{booktabs}
\usepackage{subcaption}
\usepackage{adjustbox}
\usepackage{graphicx}
\usepackage{algorithm}  
\usepackage{algpseudocode}  
\usepackage{multirow}
\usepackage[justification=justified, skip=2pt]{caption}

\usepackage[inline]{enumitem}
\usepackage[capitalize]{cleveref}  %

\usepackage{tikz}
\usetikzlibrary{arrows.meta}
\usetikzlibrary{backgrounds}
\usetikzlibrary{positioning, fit, mindmap, trees, calc, tikzmark, shapes}
\usetikzlibrary{shapes.arrows, fadings, automata, tikzmark, decorations.pathreplacing, patterns}
\usetikzlibrary{arrows.meta}
\usetikzlibrary{intersections}
\usepackage{tkz-euclide}

\usepackage{pgfplots}
\usepgfplotslibrary{groupplots}
\pgfplotsset{compat=1.14}

\usepackage{sistyle}
\SIthousandsep{,}

\usepackage{soul}
\setstcolor{red}

\usepackage{xcolor}

\usepackage{balance}

\newcommand\blfootnote[1]{%
\begingroup
\renewcommand\thefootnote{}\footnote{#1}%
\addtocounter{footnote}{-1}%
\endgroup
}

\newcommand{\ie}{i.e., }
\newcommand{\eg}{e.g., }

\definecolor{tea_green}{RGB}{214, 234, 193}
\definecolor{hint_green}{RGB}{226,246,209}
\definecolor{Madang}{RGB}{190,235,159}
\definecolor{yellow_green}{RGB}{198,222,119}
\definecolor{link_water}{RGB}{221, 232, 250}
\definecolor{celestial_blue}{RGB}{52, 152, 219}
\definecolor{shakespeare}{RGB}{85, 154, 193}
\definecolor{buttermilk}{RGB}{255,242,174}
\definecolor{chardonnay}{RGB}{250,196,114}
\definecolor{rajah}{RGB}{253,180,98}
\definecolor{fog}{RGB}{213, 193, 234}
\definecolor{melon}{RGB}{254,191,181}
\definecolor{sundown}{RGB}{249, 180, 181}
\definecolor{mona_lisa}{RGB}{246,152,134}
\definecolor{salmon}{RGB}{242,131,107}

\definecolor{saltpan}{RGB}{238, 243, 232}
\definecolor{aqua_spring}{RGB}{232, 243, 232}
\definecolor{tea_green}{RGB}{214, 234, 193}
\definecolor{Madang}{RGB}{190,235,159}
\definecolor{fringy_flower}{RGB}{194, 234, 193}
\definecolor{aero_blue}{RGB}{193, 234, 213}
\definecolor{pixie_green}{RGB}{183,214,170}
\definecolor{french_pass}{RGB}{195,232,246}
\definecolor{ice_cold}{RGB}{169,232,220}
\definecolor{pale_turquoise}{RGB}{172,240,242}
\definecolor{cruise}{RGB}{179,226,205}
\definecolor{sail}{RGB}{163,205,235}
\definecolor{spindle}{RGB}{179,205,227}
\definecolor{link_water}{RGB}{221, 232, 250}
\definecolor{periwinkle}{RGB}{203,213,232}
\definecolor{zanah}{RGB}{220, 233, 213}
\definecolor{frostee}{RGB}{217, 231, 214}
\definecolor{opal}{RGB}{199, 221, 211}
\definecolor{jet_stream}{RGB}{188, 214, 210}
\definecolor{skeptic}{RGB}{153, 187, 167}
\definecolor{hint_green}{RGB}{226,246,209}
\definecolor{snow_flurry}{RGB}{230,245,201}
\definecolor{surf_crest}{RGB}{205,230,208}
\definecolor{yellow_green}{RGB}{198,222,119}
\definecolor{cream}{RGB}{255,255,204}
\definecolor{pale_prim}{RGB}{255,255,179}
\definecolor{spring_sun}{RGB}{242,243,195}
\definecolor{portafino}{RGB}{245,237,160}
\definecolor{buttermilk}{RGB}{255,242,174}
\definecolor{cream_brulee}{RGB}{255, 229, 151}
\definecolor{dairy_cream}{RGB}{254,226,189}
\definecolor{champagne}{RGB}{254,217,166}
\definecolor{chardonnay}{RGB}{250,196,114}
\definecolor{manhattan}{RGB}{226,180,125}
\definecolor{rajah}{RGB}{253,180,98}
\definecolor{early_dawn}{RGB}{252,243,218}
\definecolor{egg_shell}{RGB}{238, 234, 215}
\definecolor{selago}{RGB}{243, 232, 243}
\definecolor{quartz}{RGB}{219,223,238}
\definecolor{fog}{RGB}{213, 193, 234}
\definecolor{languid_lavender}{RGB}{222,203,228}
\definecolor{watusi}{RGB}{254,221,207}
\definecolor{coral_andy}{RGB}{243,204,205}
\definecolor{cosmos}{RGB}{248,209,210}
\definecolor{melon}{RGB}{254,191,181}
\definecolor{azalea}{RGB}{234, 193, 194}
\definecolor{beauty_bush}{RGB}{235, 185, 179}
\definecolor{sundown}{RGB}{249, 180, 181}
\definecolor{mona_lisa}{RGB}{246,152,134}
\definecolor{salmon}{RGB}{242,131,107}

\definecolor{summer_sky}{RGB}{58, 151, 233}
\definecolor{chateau_green}{RGB}{72, 179, 96}
\definecolor{matisse}{RGB}{25, 104, 167}
\definecolor{allports}{RGB}{31, 106, 125}
\definecolor{sun_shade}{RGB}{255, 144, 68}
\definecolor{flamingo}{RGB}{237, 88, 85}
\definecolor{studio}{RGB}{128, 91, 160}

\definecolor{maya_blue}{RGB}{102, 204, 255}
\definecolor{feijoa}{RGB}{178,223,138}
\definecolor{sushi}{RGB}{117, 168, 47}
\definecolor{norway}{RGB}{158, 194, 132}
\definecolor{japanese_laurel}{RGB}{53, 116, 40}
\definecolor{see_green}{RGB}{161,228,195}
\definecolor{monte_carlo}{RGB}{135,204,194}
\definecolor{granny_smith_apple}{RGB}{150,214,150}
\definecolor{moss_green}{RGB}{170,216,176}
\definecolor{chateau_green}{RGB}{72, 179, 96}
\definecolor{opal}{RGB}{164,207,190}
\definecolor{acapulco}{RGB}{117, 170, 148}
\definecolor{viridian}{RGB}{55, 137, 122}
\definecolor{amazon}{RGB}{56, 123, 84}
\definecolor{asparagus}{RGB}{123, 160, 91}
\definecolor{fruit_salad}{RGB}{91, 160, 94}
\definecolor{puerto_rico}{RGB}{72, 179, 150}
\definecolor{mountain_meadow}{RGB}{0, 163, 136}
\definecolor{matisse}{RGB}{25, 104, 167}
\definecolor{allports}{RGB}{31, 106, 125}
\definecolor{astral}{RGB}{55, 111, 137}
\definecolor{spring_leaves}{RGB}{46, 83, 117}
\definecolor{biscay}{RGB}{44, 62, 80}
\definecolor{midnight}{RGB}{0, 29, 50}
\definecolor{amethyst}{RGB}{153, 102, 204}
\definecolor{studio}{RGB}{128, 91, 160}
\definecolor{tapestry}{RGB}{194, 109, 132}
\definecolor{atomic_tangerine}{RGB}{255, 153, 102}
\definecolor{amber}{RGB}{255, 191, 0}
\definecolor{casablanca}{RGB}{244, 178, 84}
\definecolor{california}{RGB}{233, 140, 58}
\definecolor{tomato}{RGB}{255, 97, 56} 
\definecolor{alizarin}{RGB}{233, 58, 64}

\definecolor{linen}{RGB}{251, 239, 227}
\definecolor{double_pearl_lusta}{RGB}{253, 242, 208}
\definecolor{oasis}{RGB}{253, 242, 208}
\definecolor{milan}{RGB}{255, 254, 169}
\definecolor{texas}{RGB}{245, 232, 123}
\definecolor{maize}{RGB}{249, 212, 156}

\definecolor{turmeric}{RGB}{211, 178, 76}
\definecolor{saffron}{RGB}{249,193,62}
\definecolor{my_sin}{RGB}{255, 176, 59}
\definecolor{tree_poppy}{RGB}{246, 154, 27}
\definecolor{jaffa}{RGB}{240, 131, 58}
\definecolor{crusta}{RGB}{254, 127, 44}
\definecolor{tahiti_gold}{RGB}{223, 102, 36}
\definecolor{outrageous_orange}{RGB}{255, 100, 45}
\definecolor{safety_orange}{RGB}{254, 106, 0}

\definecolor{azalea}{RGB}{251, 196, 196}
\definecolor{oyster_pink}{RGB}{238,206,205} 
\definecolor{coral_candy}{RGB}{242,208,205} 
\definecolor{baby_pink}{RGB}{246, 194, 192}
\definecolor{petite_orchid}{RGB}{223, 157, 155}
\definecolor{apricot}{RGB}{241,140,122}
\definecolor{NY_pink}{RGB}{228,136,113}
\definecolor{carmine_pink}{RGB}{231, 76, 60}
\definecolor{deep_carmine_pink}{RGB}{236, 50, 67}

\definecolor{wewak}{RGB}{244, 143, 150}
\definecolor{light_coral}{RGB}{244, 127, 123}
\definecolor{bittersweet}{RGB}{255,111,105}
\definecolor{carnation}{RGB}{245, 80, 86}
\definecolor{flamingo}{RGB}{237, 88, 85}
\definecolor{sunset_orange}{RGB}{242,89,75}
\definecolor{ku_crimson}{RGB}{243, 0, 25}
\definecolor{amaranth}{RGB}{234,46,73}
\definecolor{valencia}{RGB}{214, 87, 70}
\definecolor{chilean_fire}{RGB}{215, 87, 44}
\definecolor{mexican_red}{RGB}{170, 41, 37}

\definecolor{napa}{RGB}{163, 154, 137}

\definecolor{athens_gray}{RGB}{236, 240, 241}
\definecolor{gallery}{RGB}{240,240,240}
\definecolor{mercury}{RGB}{230,230,230}
\definecolor{platinum}{RGB}{228,228,228}
\definecolor{silver}{RGB}{191,191,191}
\definecolor{aluminum}{RGB}{153,153,153}
\definecolor{ship_gray}{RGB}{77,77,77}
\definecolor{tuatara}{RGB}{67, 67, 67}

\definecolor{malibu}{RGB}{110, 180, 240}
\definecolor{celestial_blue}{RGB}{52, 152, 219}
\definecolor{curious_blue}{RGB}{41, 128, 185}
\definecolor{french_blue}{RGB}{0, 112, 182}
\definecolor{matisse}{RGB}{25, 104, 167}
\definecolor{shakespeare}{RGB}{85, 154, 193}
\definecolor{seagull}{RGB}{128,177,211}
\definecolor{jelly_bean}{RGB}{45, 126, 150}
\definecolor{venice_blue}{RGB}{87, 135, 105}
\definecolor{boston_blue}{RGB}{68, 147, 161}

\definecolor{turquoise}{RGB}{41,217,194}
\definecolor{java}{RGB}{2,190,196}
\definecolor{riptide}{RGB}{141,211,199}
\definecolor{mountain_meadow}{RGB}{0, 163, 136}
\definecolor{free_speech_aquamarine}{RGB}{0, 156, 114}

\definecolor{cosmic_latte}{RGB}{222, 247, 229}
\definecolor{chinook}{RGB}{163, 232, 178}
\definecolor{padua}{RGB}{121, 189, 143}
\definecolor{ocean_green}{RGB}{79, 176, 112}
\definecolor{pastel_green}{RGB}{107, 227, 135}
\definecolor{chateau_green}{RGB}{69, 191, 85}
\definecolor{RoyalBlue}{RGB}{69, 191, 85}
\definecolor{pigment_green}{RGB}{0, 175, 79}
\definecolor{fern}{RGB}{101,197,117}
\definecolor{killarney}{RGB}{56, 113, 66}

\AtBeginDocument{%
  \providecommand\BibTeX{{%
    \normalfont B\kern-0.5em{\scshape i\kern-0.25em b}\kern-0.8em\TeX}}}

\renewcommand{\arraystretch}{0.93}



\begin{document}

\copyrightyear{2022}
\acmYear{2022}
\setcopyright{acmcopyright}\acmConference[WWW '22]{Proceedings of the ACM Web Conference 2022}{April 25--29, 2022}{Virtual Event, Lyon, France} \acmBooktitle{Proceedings of the ACM Web Conference 2022 (WWW '22), April 25--29, 2022, Virtual Event, Lyon, France}
\acmPrice{15.00}
\acmDOI{10.1145/3485447.3511997}
\acmISBN{978-1-4503-9096-5/22/04}

\title{Recommendation Unlearning}
\author{Chong Chen$^{1*}$, Fei Sun$^{2\dagger}$, Min Zhang$^{1\dagger}$,  Bolin Ding$^{2}$}
\affiliation{
\institution{$^1$Department of Computer Science and Technology, Institute for Artificial Intelligence,\\
Beijing National Research Center for Information Science and Technology, 
Tsinghua University, Beijing 100084, China\\
$^2$DAMO Academy, Alibaba Group}
\country{}
{cc17@mails.tsinghua.edu.cn, z-m@tsinghua.edu.cn, $\quad$ $^2$\{ofey.sf, bolin.ding\}@alibaba-inc.com}
}

\renewcommand{\authors}{Chong Chen, Fei Sun, Min Zhang, and Bolin Ding}
\renewcommand{\shortauthors}{Chong Chen, Fei Sun, Min Zhang, and Bolin Ding}

\begin{abstract}

Recommender systems provide essential web services by learning users' personal preferences from collected data. 
However, in many cases, systems also need to forget some training data. From the perspective of privacy, users desire a tool to erase the impacts of their sensitive data from the trained models.
From the perspective of utility, if a system's utility is damaged by some bad data, the system needs to forget such data to regain utility.
While unlearning is very important, it has not been well-considered in existing recommender systems. 
Although there are some researches have studied the problem of machine unlearning, existing methods can not be directly applied to recommendation as they are unable to consider the collaborative information.

In this paper, we propose RecEraser, a general and efficient machine unlearning framework tailored to recommendation tasks. 
The main idea of RecEraser is to divide the training set into multiple shards and train submodels with these shards. Specifically, to keep the collaborative information of the data, we first design three novel data partition algorithms to divide training data into balanced groups. We then further propose an adaptive aggregation method to improve the global model utility. Experimental results on three public benchmarks show that RecEraser can not only achieve efficient unlearning but also outperform the state-of-the-art unlearning methods in terms of model utility.
The source code can be found at \url{https://github.com/chenchongthu/Recommendation-Unlearning}

\blfootnote{$^*$This work was done when Chong Chen was an intern at Alibaba.}
\blfootnote{$^{\dagger}$Corresponding authors.} 
\end{abstract}

  


\maketitle

\section{Introduction}
\label{sec1}

Recommender systems provide personalized service for users to alleviate the information overload problem, playing a more and more important role in a wide range of applications, such as e-commerce~\cite{Cheng:Wide,Lv:CIKM19:SDM}, social media \cite{Covington:recsys16:Deep,Ying:KDD18:Graph,chen2019social,chen2019an}, and news portal \cite{okura2017embedding}. 
The key of personalized recommender systems is known as collaborative filtering \cite{he2020lightgcn,koren2009matrix,rendle2009bpr,hu2008collaborative}, which learns users' preference based on their historical records (e.g., views, clicks, and ratings). 

Once a recommender system is built, it has potentially memorized the training data. 
However, in many cases, a recommender system also needs to forget certain sensitive data and its complete lineage, which is called \textit{Recommendation Unlearning} in this paper. 
Consider privacy first, recent researches have shown that users' sensitive information could be leaked from the trained models, e.g., recommender systems \cite{zhang2021membership}, big pretrained \cite{Carlini:arxiv21:Extracting} and finetuned natural language models \cite{zanella2020analyzing}.
In such cases, users desire a tool to erase the impacts of their sensitive information from the trained models.
The second reason is utility. Nowadays, new data is collected incrementally to further refine existing models \cite{Zhang:SIGIR20:How}.
However, bad data (or called dirty data), \eg polluted data in poisoning attacks~\cite{Jagielski:SP18:Manipulating} or out-of-distribution (OOD) data~\cite{Carlini:SEC19:Secret}, will seriously degrade the performance of recommendation. 
Once these data are identified, the system needs to forget them to regain utility. Moreover, generally users' preferences are dynamic and changeable \cite{wang2020toward}. For example, a user who wants to buy a mobile phone will be happy to see recommendations about mobile phones. But after the purchase, she/he will not be interested in the recommendations of new mobile phones for a period of time. In this case, the user will want to delete certain data so that the system can provide more useful recommendations. 

The most naive unlearning method is to retrain from the original data after removing the samples that need to be forgotten. 
Unfortunately, the difficulty in applying this method lies in the expensive computational cost for large-scale data. 
Some efforts have been devoted to resolving the inefficiency issue of machine unlearning in the domains of computer vision and natural language processing \cite{bourtoule2021machine,chen2021graph,cao2015towards,ginart2019making,golatkar2020eternal}.
For example, the SISA method \cite{bourtoule2021machine} randomly split the training data into several disjoint shards and then train submodels based on each shard. After that, the final prediction results are obtained from the aggregation of submodels through majority voting or average. When some data samples are required to be forgotten, only the corresponding submodel needs to be retrained. However, existing methods cannot be directly applied to recommendation tasks. Since recommender systems rely on collaborative information across users and items, randomly dividing the data into shards could seriously damage the recommendation performance. Moreover, the aggregation part in existing unlearning methods often assign a static weight to each submodel. Although the recent method GraphEraser\cite{chen2021graph} uses a learning-based method to assign weights, the weights cannot be changed adaptively when predicting different user-item interactions. 

In light of the above problems of existing solutions, in this work we propose a novel and efficient erasable recommendation framework, namely \textit{RecEraser}, to achieve efficient unlearning while maintaining high recommendation performance. The general idea of RecEraser is to divide the training set into multiple shards and train a submodel for each shard. To keep the collaborative information of the data, we design three data partition strategies based on the similarities of users, items, and interactions, respectively. Different from traditional community detection and clustering methods \cite{girvan2002community,choi2021local,kanungo2002efficient}, our data partition strategies are designed to achieve balanced partition so that the unlearning efficiency will not be affected by the unbalanced shard size. In addition, considering that recommender systems usually face a variety of users and items, submodels should have different contributions for the prediction of different user-item pairs. To further improves the recommendation performance, we propose an attention-based adaptive aggregation method.
To evaluate our method, we apply RecEraser on three real-world datasets with extensive experiments. Since the architecture of RecEraser is model-agnostic for base models, we utilize three representative recommendation models BPR \cite{rendle2009bpr}, WMF \cite{hu2008collaborative,chen2020efficient2}, and LightGCN \cite{he2020lightgcn} as its base models. The experimental results show that RecEraser can not only achieve efficient unlearning, but also outperform the state-of-the-art unlearning frameworks like SISA \cite{bourtoule2021machine} and GraphEraser \cite{chen2021graph} in terms of performance. 
Further ablation studies also show the effectiveness of our proposed data partition strategies and adaptive aggregation method. The main contributions of this work are:
\begin{enumerate}[leftmargin=0.5cm]
  \item To the best of our knowledge, this is the first work that addresses the machine unlearning problem for recommendation. A general erasable recommendation framework named RecEraser is proposed to achieve both high unlearning efficiency and high performance.
  \item We design three data partition strategies to split recommendation data into balanced shards, and propose an attention-based adaptive aggregation method to further improve the performance of RecEraser.
  \item We conduct extensive experiments on three real-world datasets and three representative recommendation models. The results show that RecEraser can not only achieve efficient unlearning, but also outperform the state-of-the-art unlearning framework in terms of recommendation performance. 
\end{enumerate}
\section{Related Work}

\subsection{Item Recommendation}

Popularized by the Netflix Challenge, early recommendation methods \cite{koren2009matrix,koren2008factorization} were designed to model users' explicit feedback by mapping users and items to a latent factor space like matrix factorization. Later on, researchers found that users interact with items mainly through implicit feedback, such as purchases on E-commerce sites and watches on online video platforms. Then a surge of recommendation methods were proposed for learning from implicit feedback \cite{hu2008collaborative,rendle2009bpr,he2017neural,chen2020efficient3,chen2020efficient}. 
Specifically, \citet{hu2008collaborative} proposed a non-sampling based method WMF, which assumes that all unobserved items are negative samples. Some recent studies have also been made to resolving the inefficiency issue of non-sampling learning. For instance, \citet{chen2020efficient2} derive a flexible non-sampling loss for recommendation models, which achieves both effective and efficient performance. On another line of research, \citet{rendle2009bpr} proposed a pair-wise learning method BPR, which is a sampling-based method that optimizes the model based on the relative preference of a user over pairs of items. 

Due to the popularity of deep learning, there is a large literature exploiting different neural networks for recommender systems. 
In~\cite{he2017neural}, He et al.~presented a Neural Collaborative Filtering (NCF) framework to address implicit feedback data by jointly learning a matrix factorization and a feedforward neural network. 
The NCF framework has been widely extended to adapt to different recommendation scenarios \cite{wang2017item,he2017neural2}. Recently, it has become a trend to explore the application of newly proposed deep learning architectures in recommendation. Such as attention mechanisms \cite{xue2019deep,chen2018neural}, Convolutional Neural Network \cite{he2018outer,zheng2017joint}, Recurrent Neural Network \cite{okura2017embedding}, and Graph Neural Networks \cite{fan2019graph,wang2019neural,chen2021graph2}. 
Specifically, \citet{wang2019neural} proposed NGCF to exploit high-order proximity by propagating embeddings on the user-item interaction graph. NGCF is then further extended to LightGCN \cite{he2020lightgcn} by removing the non-linear activation function and feature transformation. LightGCN is more efficient than vanilla GCN modes and has achieved state-of-the-art performance in Top-$K$ recommendation task.

\subsection{Machine Unlearning}

Machine unlearning\footnote{It is worth noting that the purpose of unlearning is different from differential privacy (DP) methods \cite{dwork2014algorithmic,gao2020dplcf} which aim to protect users' privacy information instead of deleting them.
Besides, unlearning usually requires more strict guarantee than DP.}, also known as \textit{selective forgetting}~\cite{golatkar2020eternal} or \textit{data removal/deletion}~\cite{Guo:ICML20:Certified,Ginart:NIPS19:Making} in machine learning, refers to a process that aims to remove the influence of a specified subset of training data upon request from a trained model.
Previous studies on machine unlearning can be divided into two groups: approximate unlearning and exact unlearning.

\textbf{Approximate unlearning} offers statistical guarantee about data deletion, thus a.k.a statistical unlearning~\cite{Guo:ICML20:Certified,golatkar2020eternal,Izzo:AISTATS21:Approximate}.
The essential idea is to relax the requirements for exact deletion, \ie it only provides a statistical guarantee that an unlearned model cannot be distinguished from a model that was never trained on the removed data~\cite{Guo:ICML20:Certified}. 
They usually employ gradient based update strategies to quickly eliminate the influence of samples that are requested to be deleted~\cite{Neel:ALT21:Descent}.
For example, \citet{Guo:ICML20:Certified}, \citet{golatkar2020eternal}, and \citet{Golatkar:ECCV20:Forgetting} proposed different Newton's methods to approximate retraining for convex models, \eg linear regression, logistic regression, and the last fully connected layer of a neural network.
An alternative is to eliminate the influence of the samples that need to be deleted to the learned model based on influence functions~\cite{Izzo:AISTATS21:Approximate}.
Compared with exact unlearning, approximate unlearning methods are usually more efficient.
However, their guarantees are probabilistic and are hard to apply on non-convex models like deep neural networks.
It makes them not very suitable for the applications of recommender systems, which are strictly regulated by the laws, \eg GDPR and CCPA.

\textbf{Exact unlearning} aims to ensure that the request data is completely removed from the learned model.
Early works usually aim to speed up the exact unlearning for simple models or under some specific conditions~\cite{Cauwenberghs:NIPS00:Incremental,cao2015towards,Schelter2020AmnesiaM}, like leave-one-out cross-validation for SVMs (Support Vector Machines)~\cite{Cauwenberghs:NIPS00:Incremental,Karasuyama:NIPS09:Multiple}, provably efficient data deletion in $k$-means clustering~\cite{Ginart:NIPS19:Making}, and
fast data deletion for Na\"ive Bayes based on statistical query learning which assumes the training data is in a decided order~\cite{cao2015towards}.
More recently, the representative work is SISA (Sharded, Isolated, Sliced, and Aggregated)~\cite{bourtoule2021machine}.
SISA is a quite general framework, and its key idea can be abstracted into three steps: 
\begin{enumerate*}
    \item divide the training data into several disjoint shards;
    \item train sub-models independently (\ie without communication) on each of these data shards;
    \item aggregate the results from all shards for final prediction.
\end{enumerate*}
In this way, unlearning can be effectively achieved by only retraining the affected sub-model.
Subsequently, \citet{chen2021graph} applied this idea to the unlearning of graph with an improved sharding algorithm. Our RecEraser is different from existing methods in the following aspects: (1) we design new data partition methods to keep the collaborative information of the data; (2) we propose an adaptive aggregation method to improve the global model utility. These designs make our RecEraser more suitable for recommendation tasks.

\section{Recommendation Unlearning}
We first introduce the key notation, then formulate the recommendation unlearning task and discuss its differences and challenges with existing methods.

\subsection{Notations and Problem Formulation}

\begin{table}[]
\caption{Summary of symbols and notations.}
\label{table1}
\begin{adjustbox}{max width = \textwidth}
\begin{tabular}{lp{6.3cm}}
\toprule
\textbf{Symbol}                & \textbf{Description} \\ \midrule
$\textbf{U}, \textbf{V}$                     & Set of users and items, respectively                                             \\

$\textbf{Y}$                     & User-item interaction matrix                                  \\
$K$                     & The number of shards                                   \\
$S_i$                     & Shards $i$                                  \\
$M_i$  &The submodel trained on $S_i$
\\
$\bar{\textbf{p}}_u, \bar{\textbf{q}}_v$                   & The pre-trained embeddings of user $u$ and item $v$, respectively, used for data partition  \\

$\textbf{p}_u^i, \textbf{q}_v^i$                   & The embeddings of user $u$ and item $v$ learned by $M_i$, respectively  \\

$\textbf{p}_u, \textbf{q}_v$                   & The aggregated embeddings of user $u$ and item $v$, respectively  \\

\bottomrule
\end{tabular}
\end{adjustbox}
\end{table}

Table \ref{table1} depicts the notations and key concepts used in this paper. We denote the user and item sets as $\textbf{U}$ and $\textbf{V}$, respectively. The user-item interaction matrix is denoted as $\textbf {Y}=[y_{uv}] \in \{0,1\}$, indicating whether $u$ has an interaction with item $v$. Given a target user $u$, the recommendation task is to recommend a list of items that $u$ may be interested in.

For \textit{Recommendation Unlearning}, 
if user $u$ wants to revoke one record to item $v$ (i.e., $y_{uv}$), a recommender system needs to obtain an unlearned model trained on $\textbf{Y} \backslash y_{uv}$.
Formally, the task of recommended unlearning is to achieve three general objectives.
\begin{itemize}
    \item \textbf{Provable Guarantees}: It is the basic requirement of unlearning which demands the revoked data must be really unlearned and do not influence model parameters.
    \item \textbf{High Unlearning Efficiency}: The Unlearning process of forgetting the required samples should be as fast as possible.
    \item \textbf{Comparable Performance}: The performance of the unlearned model's prediction should be comparable to that of retraining from scratch.
\end{itemize}

\subsection{Challenges of Recommendation Unlearning}

Existing machine unlearning methods are mainly designed in the domains of computer vision and natural language processing, which can not be directly applied to recommendation tasks. For example, the state-of-the-art unlearning method SISA \cite{bourtoule2021machine} uses a random method to divide the training data into multiple shards. However, the inputs of recommendation are user-item interactions, which contain rich collaborative information. Randomly divide the training data will destroy the collaborative information which could damage the recommendation performance. One promising approach is using community detection or clustering methods \cite{girvan2002community,choi2021local,kanungo2002efficient}. However, due to the underlying structure of real-world data, these methods may lead to highly unbalanced data partition. Therefore, the unlearning efficiency will be affected if the required samples belongs to a large shard. 
To address the above problems, we need to design new balanced data partition methods that can preserve the collaborative information between users and items. Moreover, existing unlearning methods typically use a static aggregation strategy at the inference phase, which is not suitable for recommender systems that usually face a variety of users and items. Although the recent method GraphEraser\cite{chen2021graph} uses a learning-based method to assign weights, the weights cannot be changed adaptively when predicting different user-item interactions. 
To further improves the performance of a recommendation unlearning framework, the weight of a submodel should depend on how well it learns the feature of an individual user (or an item).

\section{RecEraser Method}

In this section, we first present a general overview of the RecEraser framework, then introduce the two key ingredients of our proposed model in detail, which are: 1) balanced data partition and 2) attention-based adaptive aggregation.

\begin{figure}
\includegraphics[width=3in]{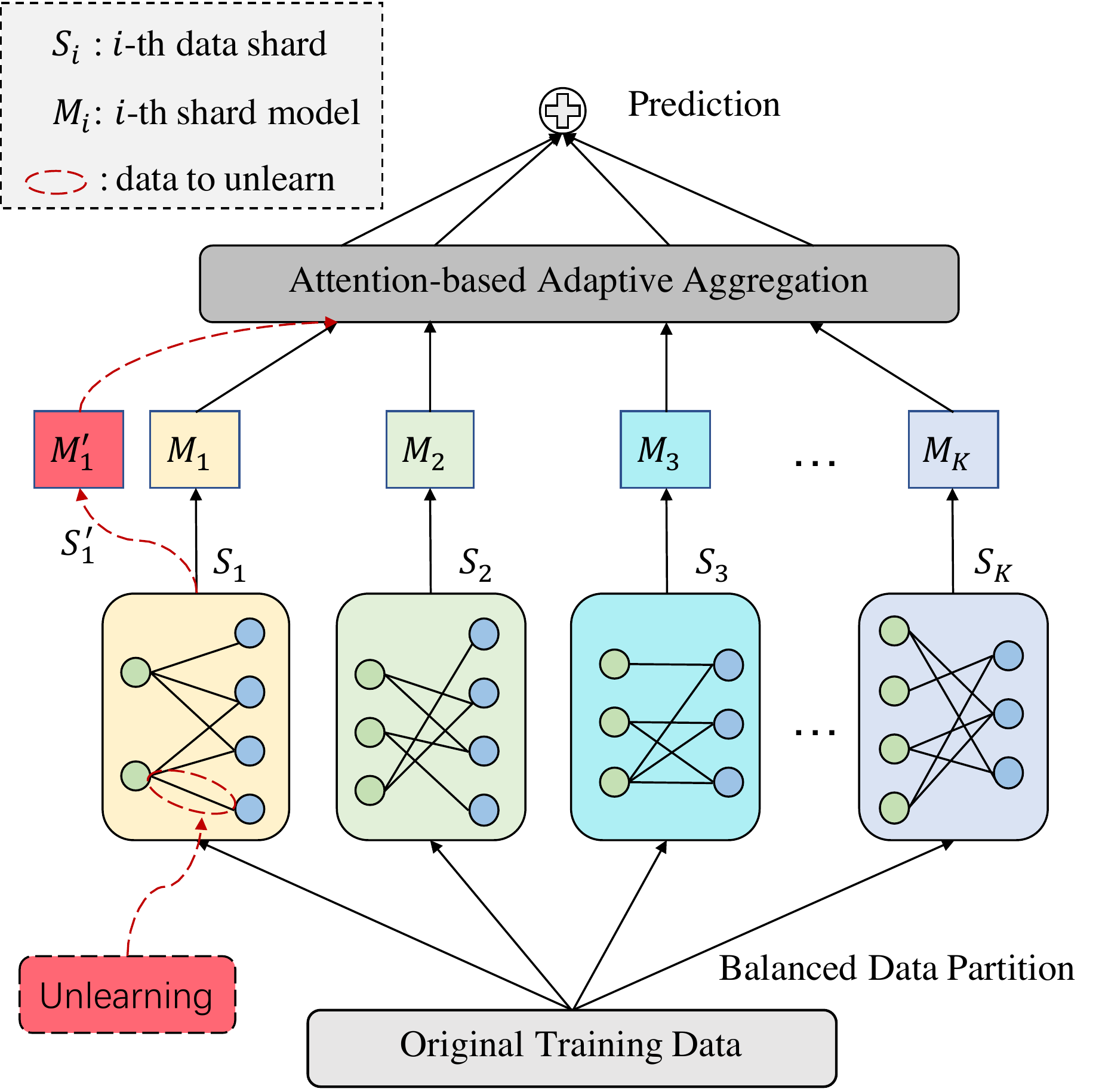}
\centering
\caption{Illustration of our RecEraser framework, which consists of three parts: data partition, submodel training, and submodel aggregation. When receiving an unlearning request of data, only the corresponding submodel and the aggregation part need to be retrained.}
\label{figure1}
\end{figure}

\subsection{Overview}

The overall framework of RecEraser is described in Figure \ref{figure1}. From the figure, we first make a simple overview of our method:

\begin{enumerate}[leftmargin=0.5cm]
 \item RecEraser consists of three phases: balanced data partition, submodel training, and attention-based adaptive aggregation.

\item The balanced data partition part is designed to divide the training data while preserving the collaborative information. After the data is partitioned, a submodel ($M_i$) is trained for each of the shard data ($S_i$). All submodels share the same model architecture and can be trained in parallel to speed up the training process. At the prediction phase, for each individual prediction, an attention-based adaptive aggregation strategy is applied to find the optimal weight of different submodels. 

\item When data needs to be unlearned, only one of the submodels whose shard contains the point to be unlearned and the aggregation part need to be retrained, which is much more efficient than retraining the whole model from scratch.
\end{enumerate}

\subsection{Balanced Data Partition}

As we have mentioned before, the data used for recommendation tasks usually contains rich collaborative information. To preserve the collaborative information, one promising approach is to rely on community detection and clustering methods \cite{girvan2002community,choi2021local,kanungo2002efficient}. However, direct application of them may lead to highly unbalanced data partition.
Motivated by \cite{chen2021graph}, we propose three new balanced data partition methods, namely User-based Balanced Partition (UBP), Item-based Balanced Partition (IBP), and Interaction-based Balanced Partition (InBP). Unlike \cite{chen2021graph} that divides data through the k-means of all node embeddings, our three data partition strategies are based on the similarities of users, items, and interactions, respectively, which is more flexible for recommendation tasks.

\begin{algorithm}[t]  
  \caption{User-based Balanced Partition algorithm (UBP)}  
  \label{alg1}  
  \begin{algorithmic}[1]  
    \Require  
      Pre-trained user embeddings $\bar{\textbf{P}}$= $\{\bar{\textbf{p}}_1,\bar{\textbf{p}}_2,\dots,\bar{\textbf{p}}_m\}$; user-item interactions \textbf{Y}; number of shards $K$; maximum number of each shard $t$
    \Ensure Shards 
    \textbf{S}= $\{S_1,S_2,\dots,S_K\}$
    \State Randomly select $K$ anchors \textbf{A}= $\{a_1,a_2,\dots,a_K\}$ from \textbf{U}
    \While{\textit{Stopping criteria is not met}} 
    
    \For{each $a_i$ in \textbf{A}}
    \For{each $u$ in \textbf{U}}
     \State 
     Calculate the distance $E(a_i,u)=\mathrm{dist}(a_i,u)$ (Eq.(\ref{eq1}))
     \EndFor
     \EndFor
      \State Sort $E$ in ascending order to get $E_s$
\State Empty \textbf{S}
\For {each $a_i$ and $u$ in $E_s$}
\If{$|S_i|<t$ and $u$ has not been assigned}

     \State $S_i \leftarrow S_i \cup \textbf{Y}_u$
     \EndIf
     \EndFor
     \State Updating \textbf{A} by Eq.(\ref{eq2})
    \EndWhile \\
    \Return \textbf{S}
  \end{algorithmic} 
\end{algorithm} 
\begin{algorithm}[t]  
  \caption{Interaction-based Balanced Partition algorithm (InBP)}  
  \label{alg2}  
  \begin{algorithmic}[1]  
    \Require  
      Pre-trained user embeddings $\bar{\textbf{P}}$= $\{\bar{\textbf{p}}_1,\bar{\textbf{p}}_2,\dots,\bar{\textbf{p}}_m\}$; item embeddings $\bar{\textbf{Q}}$= $\{\bar{\textbf{q}}_1,\bar{\textbf{q}}_2,\dots,\bar{\textbf{q}}_n\}$; user-item interactions \textbf{Y}; number of shards $K$; maximum number of each shard $t$
    \Ensure Shards 
    \textbf{S}= $\{S_1,S_2,\dots,S_K\}$
    \State Randomly select $K$ anchors \textbf{A}= $\{a_1,a_2,\dots,a_K\}$ from \textbf{Y}
    \While{\textit{Stopping criteria is not met}} 
    
    \For{each $a_i$ in \textbf{A}}
    \For{each $y_{uv}$ in \textbf{Y}}
     \State 
     Calculate $E(a_i,y_{uv})=\mathrm{dist}(a_i,y_{uv})$ (Eq.(\ref{eq3}))
     \EndFor
     \EndFor
      \State Sort $E$ in ascending order to get $E_s$
\State Empty \textbf{S}
\For {each $a_i$ and $y_{uv}$ in $E_s$}
\If{$|S_i|<t$ and $y_{uv}$ has not been assigned}
     \State $S_i \leftarrow S_i \cup y_{uv}$
     \EndIf
     \EndFor
     \State Updating \textbf{A} by Eq.(\ref{eq4})
    \EndWhile \\
    \Return \textbf{S}
  \end{algorithmic} 
\end{algorithm} 

The UBP method divides training data based on the similarities of users, which is shown in Algorithm \ref{alg1}. 
Given an entire user set \textbf{U}, $K$ anchor users are first randomly selected as the centers of each submodel, denoted by \textbf{A}= $\{a_1,a_2,\dots,a_K\}$. For each anchor user, the corresponding shard is discovered by estimating the distance with respect to all other users. Following the setting of $k$-means \cite{kanungo2002efficient}, the Euclidean distance is employed in this work, which is defined as follows: 
\begin{equation}
\begin{split}
\label{eq1}
\mathrm{dist}(a_i,u)=\left\|\bar{\textbf{p}}_{a_i}-\bar{\textbf{p}}_{u}\right\|_{2}= \sqrt{\sum_{j=1}^{n}\left(\bar{p}_{a_i,j}-\bar{p}_{u,j}\right)^{2}}
\end{split}
\end{equation}
where $\bar{\textbf{p}}_{a_i}$ and $\bar{\textbf{p}}_{u}$ are the embedding vectors for the anchor user $a_i$ and the user $u$. The embedding vectors are computed by a pre-trained model (WMF \cite{hu2008collaborative} is used in this work, but any other embedding vectors can be applied).

After that, we get $m\times K$ distance scores, where $m$ denotes the number of users. These scores are sorted in ascending order to get $E_s$. For each user-anchor pair in $E_s$, if the size of $S_i$ is smaller than the maximum number of shard $t$, then the training interaction $\textbf{Y}_u$ is allocated to shard $S_i$. Finally, the new anchors are calculated as the average of all the users in their corresponding shards:
\begin{equation}
\begin{split}
\label{eq2}
a_i=\frac{\sum_{j \in S_i} \bar{\textbf{p}}_{j}}{\left| S_i\right|}
\end{split}
\end{equation}
The UBP method repeats the above steps until it meets the stopping criteria (reaches the maximum iteration or the anchors do not change).
The IBP method is based on the similarities of pre-trained item embeddings, which is similar to UBP with little variations. To avoid repetition, we do not introduce it step by step.

UBP and IBP use users and items to model the local property respectively, which may ignore the global feature of training interaction (user-item pairs). So we further propose an \textbf{In}teraction-based \textbf{B}alanced \textbf{P}artition method (InBP), which considers the similarities of user-item pairs to divide the training data. As shown in Algorithm \ref{alg2}, the inputs of InBP include the pre-trained user embeddings $\bar{\textbf{P}}$= $\{\bar{\textbf{p}}_1,\bar{\textbf{p}}_2,\dots,\bar{\textbf{p}}_m\}$, item embeddings $\bar{\textbf{Q}}$= $\{\bar{\textbf{q}}_1,\bar{\textbf{q}}_2,\dots,\bar{\textbf{q}}_n\}$, user-item interactions \textbf{Y}, the number of shards $K$, and the maximum number of shard $t$. The main differences between InBP and UBP is as follows:
(1) InBP randomly select $K$ anchors \textbf{A}= $\{a_1,a_2,\dots,a_K\}$ from user-item interactions \textbf{Y}, where $a_i=(\textbf{p}_i,\textbf{q}_i)$, $\textbf{p}_{i}$ and $\textbf{q}_{i}$ are the embedding vectors for the data point $a_i$;
(2) The distance between anchor $a_i$ and interaction $y_{uv}$ is calculated as follows:
\begin{equation}
\begin{split}
\label{eq3}
\mathrm{dist}(a_i,y_{uv})&=\left\|\bar{\textbf{p}}_{i}-\bar{\textbf{p}}_{u}\right\|_{2} \times \left\|\bar{\textbf{q}}_{i}-\bar{\textbf{q}}_{v}\right\|_{2}\\
&=\sqrt{\sum_{j=1}^{n}\left(\bar{p}_{i,j}-\bar{p}_{u,j}\right)^{2}} \times \sqrt{\sum_{j=1}^{n}\left(\bar{q}_{i,j}-\bar{q}_{v,j}\right)^{2}}
\end{split}
\end{equation}
where $\bar{\textbf{p}}_{u}$ and $\bar{\textbf{q}_{v}}$ are the pre-trained embedding vectors for the data point $y_{uv}$;
(3) For each interaction pair in $E_s$, if the size of $S_i$ is smaller than the maximum number of shard $t$, then the training interaction $y_{uv}$ is allocated to shard $S_i$; (4) The new anchors are updated as follows:
\begin{equation}
\begin{split}
\label{eq4}
a_i=\left(\frac{\sum_{j \in S_i} \bar{\textbf{p}}_{j}}{\left| S_i\right|}, \frac{\sum_{j \in S_i} \bar{\textbf{q}}_{j}}{\left| S_i\right|}\right)
\end{split}
\end{equation}

\subsection{Attention-based Adaptive Aggregation}

After the training data is partitioned, a submodel $M_i$ is trained on each shard data $S_i$. Generally, various submodels should have different contributions for the prediction of different user-item pairs. For example, if shard $S_i$ contains more training interactions of user $u$ than shard $S_j$, then the weight of $M_i$ is more likely to be greater than that of $M_j$ when predicting $u$'s preference. Although the recent method GraphEraser\cite{chen2021graph} uses a learning-based method to assign weights, the weights cannot be changed when predicting different user-item interactions. 
To further improves the recommendation performance, we propose an attention-based adaptive aggregation method.

Considering that the representations of users and items learned by different submodels may be embedded in different spaces, we first transfer them into the same representation space. Specifically, for $\textbf{P}^i$ and $\textbf{Q}^i$ learned by submodel $S_i$, the transfer scheme is:
\begin{equation}
\begin{split}
\label{eq5}
\textbf{P}^i_{\mathit{tr}}=\textbf{W}^i\textbf{P}^i+\textbf{b}^i; \quad    \textbf{Q}^i_{\mathit{tr}}=\textbf{W}^i\textbf{Q}^i+\textbf{b}^i
\end{split}
\end{equation}
where $\textbf{W}^{i}\in \mathbb{R}^{d\times d}$ is a transfer matrix which projects $\textbf{P}^i$ and $\textbf{Q}^i$ to the global representation space, $\textbf{b}^i\in \mathbb{R}^d$ is the bias vector, and $d$ denotes the embedding size.

Then the aggregated embeddings of users and items are calculated as follows:
\begin{equation}
\begin{split}
\label{eq6}
\textbf{P}=\sum_{i=1}^K \alpha_{i} \mathbf{P}^i_{\mathit{tr}}; \quad
\textbf{Q}=\sum_{i=1}^K \beta_{i} \mathbf{Q}^i_{\mathit{tr}}
\end{split}
\end{equation}
where $\alpha_{i}$ and $\beta_{i}$ are the attention weights for aggregating the embeddings of users and items, indicating how much information being propagated from submodel $S_i$. More precisely, $\alpha_{i}$ and $\beta_{i}$ are defined as:
\begin{equation}
\begin{split}
\alpha_{i}^*=\textbf{h}_1^{\top}\sigma(\textbf{W}_1 \textbf{p}^i_{\mathit{tr}}+\textbf{b}_1);\quad \alpha_{i}=\frac{\exp(\alpha_{i}^*)}{\sum_{j=1}^K \exp(\alpha_{j}^*)}\\
\beta_{i}^*=\textbf{h}_2^{\top}\sigma(\textbf{W}_2 \textbf{p}^i_{\mathit{tr}}+\textbf{b}_2);\quad \beta_{i}=\frac{\exp(\beta_{i}^*)}{\sum_{j=1}^K \exp(\beta_{j}^*)}
\end{split}
\label{eq7}
\end{equation}
where $\textbf{W}_{1}\in \mathbb{R}^{k\times d}$, $\textbf{b}_1\in \mathbb{R}^k$, and $\textbf{h}_1\in \mathbb{R}^k$ are parameters of user attention, $\textbf{W}_{2}\in \mathbb{R}^{k\times d}$, $\textbf{b}_2\in \mathbb{R}^k$, and $\textbf{h}_2 \in \mathbb{R}^k$ are parameters of item attention. $k$ is the dimension of attention size, and $\sigma$ is the nonlinear activation function ReLU \cite{nair2010rectified}. Attention weights across all shards are normalized by the softmax function. 

The attention-based adaptive aggregation method assigns weights for different submodels based on the following optimization: 
\begin{equation}
\label{eq8}
\begin{split}
\min_\Theta \mathcal {L}(\textbf{P},\textbf{Q},\textbf{Y})+\lambda\|\Theta\|_{2}^{2}
\end{split}
\end{equation}
where \textbf{P} and \textbf{Q} are the aggregated embeddings of users and items, \textbf{Y} is the true lable of user-item interaction, $\mathcal {L}$ represents the loss function of base model (e.g., BPR pair-wise loss\cite{rendle2009bpr}), and $\Theta$ denotes the parameters of attention network. $\ell_2$ regularization parameterized by $\lambda$ on $\Theta$ is conducted to prevent overfitting. Note that only the parameters of attention network are optimized through Eq.(\ref{eq8}) while the embeddings $\textbf{P}^i$ and $\textbf{Q}^i$ learned by submodels are fixed.

The whole training data \textbf{Y} is used for training, which can also be requested to be revoked. Therefore, our aggregation method needs to be retrained when receiving an unlearning request of data, and this training time is also part of the unlearning time. Fortunately, the proposed aggregation method only needs a few epochs (about 10) to learn the optimal attention scores, and can effectively improve the performance compared to other aggregation methods.

\subsection{Training}

The architecture of our RecEraser is model-agnostic for base models. In this work we instantiate RecEraser with three representative recommendation models: BPR \cite{rendle2009bpr}, WMF \cite{hu2008collaborative}, and LightGCN \cite{he2020lightgcn}. 
The scoring functions of the above methods are all based on vector inner product. Formally, for each submodel $M_i$, the scoring function is defined as $\hat{y}_{uv}={\textbf{p}^{i}_u}^{\top}\textbf{q}_v^i$ where $\textbf{p}_u^i$ and $\textbf{q}_v^i$ denote the embeddings of user $u$ and item $v$ learned by $M_i$, respectively.

To optimize the objective function, we use mini-batch Adagrad~\cite{duchi2011adaptive} as the optimizer. Its main advantage is that the learning rate can be self-adaptive during model learning. RecEraser adopts a two-step training process. The first step is training submodels. To speed up the training process, we can also train different submodels in parallel. The second step is fixing submodels and training the attention-based aggregation part through Eq.(\ref{eq8}).

\section{Experiments}

\subsection{Experimental Setup}

\subsubsection{\textbf{Datasets}}

\begin{table}[]
\renewcommand{\arraystretch}{1.1}
\caption{Statistical details of the evaluation datasets. }
\centering
\label{table2}
\begin{tabular}{@{}lrrrrr@{}}
\toprule
\textbf{Dataset}         & \textbf{\#User} & \textbf{\#Item}  & \textbf{\#Interaction} & \textbf{Density} \\ \midrule
\textbf{Yelp2018} & 31,668           & 38,048            & 1,561,406         & 0.13\%              \\
\textbf{Movielens-1m} & 6,940           & 3,706          & 1,000,209       & 3.89\%              \\ 
\textbf{Movielens-10m} & 71,567           & 10,681                         & 10,000,054    &   1.31\%       \\ \bottomrule
\end{tabular}
\end{table}

We conduct our experiments on three real-world and publicly available datasets, which have been widely used to evaluate the performance of recommendation models: \textbf{Yelp2018\footnote{https://www.yelp.com/dataset/challenge}}, \textbf{Movielens-1m\footnote{https://grouplens.org/datasets/movielens/1m/}}, and \textbf{Movielens-10m\footnote{https://grouplens.org/datasets/movielens/10m/}}. Since we focus on the Top-$N$ recommendation task, we follow the widely used pre-processing method to convert these datasets into implicit feedbacks. Specifically, the detailed rating is transformed into a value of 0 or 1 indicating whether a user has interacted with an item. The statistical details of these datasets are summarized in Table \ref{table2}.

\begin{table*}[]
\setlength{\belowbottomsep}{0.65ex}
\caption{Comparison of different unlearning methods for recommendation unlearning. The proposed InBP method is used for data partition of RecEraser in this Table. For efficient unlearning methods SISA, GraphEraser, and our RecEraser, best results are highlighted in bold and the superscript ** indicates $p <$ 0.01 for the paired t-test of RecEraser vs. SISA and GraphEraser.}
\label{table3}
\centering
\begin{adjustbox}{max width=\textwidth}
\begin{tabular}{@{}lcccccccccccc@{}}
\toprule
\multirow{2}{*}{\textbf{Yelp2018}} & \multicolumn{4}{c}{\textbf{BPR}} & \multicolumn{4}{c}{\textbf{WMF}}& \multicolumn{4}{c}{\textbf{LightGCN}}  
\\ \cmidrule(lr){2-5} \cmidrule(lr){6-9} \cmidrule(lr){10-13}
                                 & \textbf{Retrain} & \textbf{SISA} & \textbf{GraphEraser} & \textbf{RecEraser} & \textbf{Retrain} & \textbf{SISA} & \textbf{GraphEraser} & \textbf{RecEraser} & \textbf{Retrain} & \textbf{SISA} & \textbf{GraphEraser} & \textbf{RecEraser} \\ \midrule

Recall@10            &  0.0302             &   0.0151         &   0.0224           & \textbf{0.0249**}       &   0.0368 &0.0235 &0.0250 & \textbf{0.0335**} &0.0367 &0.0221 &0.0278 &\textbf{0.0328**}     \\
Recall@20        &  0.0521              & 0.0272               &  0.0397              & \textbf{0.0439**}  &0.0619 &0.0401 &0.0431 &\textbf{0.0572**} &0.0637 &0.0392 &0.0479 &\textbf{0.0568**}                \\ 
Recall@50       &   0.1028            & 0.0566              &    0.0798            & \textbf{0.0889**}      &0.1189 &0.0785 &0.0825 &\textbf{0.1105**} &0.1216 &0.0768 &0.0924 &\textbf{0.1112**}          \\  \midrule

NDCG@10      &  0.0344              & 0.0181               &  0.0254            &  \textbf{0.0282** }   &0.0422 &0.0278 &0.0295 & \textbf{0.0385**}   &0.0424 &0.0259 &0.0319 & \textbf{0.0377**}        \\
NDCG@20       &  0.0423            &0.0224              &  0.0319              &  \textbf{0.0353**}      &0.0512 &0.0336 &0.0360 & \textbf{0.0471**}   &0.0524 &0.0321 &0.0393 &\textbf{0.0465**}       \\
NDCG@50       &   0.0612             &  0.0334            &   0.0468            &   \textbf{0.0523**}   &0.0723 &0.0478 &0.0508 &\textbf{0.0669**}   &0.0735 &0.0461 &0.0557 &\textbf{0.0669**}              \\   \bottomrule

\multirow{2}{*}{\textbf{Movielens-1m}} & \multicolumn{4}{c}{\textbf{BPR}} & \multicolumn{4}{c}{\textbf{WMF}}& \multicolumn{4}{c}{\textbf{LightGCN}}  
\\ \cmidrule(lr){2-5} \cmidrule(lr){6-9} \cmidrule(lr){10-13}
                                 & \textbf{Retrain} & \textbf{SISA} & \textbf{GraphEraser} & \textbf{RecEraser} & \textbf{Retrain} & \textbf{SISA} & \textbf{GraphEraser} & \textbf{RecEraser} & \textbf{Retrain} & \textbf{SISA} & \textbf{GraphEraser} & \textbf{RecEraser} \\ \midrule

Recall@10            &  0.1510             &   0.1080            &    0.1190             &  \textbf{0.1393**}   &0.1592 &0.0759 &0.1127 &\textbf{0.1435**} &0.1544 &0.1089 &0.1174 &\textbf{0.1425**}               \\
Recall@20        &  0.2389              & 0.1737               &  0.1906              & \textbf{0.2254**}      &0.2507 &0.1266 &0.1823 &\textbf{0.2254**} &0.2428 &0.1743 &0.1914 &\textbf{0.2266**}            \\ 
Recall@50       &   0.4005            & 0.2970               &    0.3271             & \textbf{0.3822**}        &0.4165 &0.2309 &0.3081 &\textbf{0.3818**}&0.4076 &0.3084 &0.3322 &\textbf{0.3839**}           \\  \midrule

NDCG@10      &  0.3412              & 0.2774               &  0.2844            &  \textbf{0.3208**}          &0.3546 &0.2058 &0.2575 & \textbf{0.3317**} &0.3509 &0.2767 &0.3832 &\textbf{0.3276**}           \\
NDCG@20       &  0.3368            &0.2679               &  0.2827              &   \textbf{ 0.3186**}         &0.3509 &0.1998 &0.2557 & \textbf{0.3253**} &0.3447 &0.2686 &0.2791 &\textbf{0.3229**}        \\
NDCG@50       &   0.3669             &  0.2840            &   0.3105           &   \textbf{0.3483**}            &0.3838 &0.2174 &0.2807 &\textbf{0.3542**} &0.3754 &0.2905 &0.3043 & \textbf{0.3520**}   \\   \bottomrule

\multirow{2}{*}{\textbf{Movielens-10m}} & \multicolumn{4}{c}{\textbf{BPR}} & \multicolumn{4}{c}{\textbf{WMF}}& \multicolumn{4}{c}{\textbf{LightGCN}}  
\\ \cmidrule(lr){2-5} \cmidrule(lr){6-9} \cmidrule(lr){10-13}
                                 & \textbf{Retrain} & \textbf{SISA} & \textbf{GraphEraser} & \textbf{RecEraser} & \textbf{Retrain} & \textbf{SISA} & \textbf{GraphEraser} & \textbf{RecEraser} & \textbf{Retrain} & \textbf{SISA} & \textbf{GraphEraser} & \textbf{RecEraser} \\ \midrule

Recall@10            &  0.1951             & 0.1620              &   0.1556             &  \textbf{0.1798**}    &0.2094 &0.1332 &0.1135 & \textbf{0.1994**}    &0.1994 &0.1221 & 0.1055& \textbf{0.1881**}    \\
Recall@20        &  0.3016              & 0.2507               &  0.2470              & \textbf{0.2808**}       &0.3177 &0.2142 &0.1766 & \textbf{0.3031**}   &0.3059 & 0.2033& 0.1676& \textbf{0.2899**}    \\ 
Recall@50       &   0.4688            & 0.3906               &    0.3979             & \textbf{0.4436**}        &0.4838 &0.3515 &0.2838 &\textbf{0.4655**}    &0.4748 & 0.3314&0.2632 & \textbf{0.4537** } \\  \midrule

NDCG@10      &  0.3425              & 0.3002               &  0.2688            & \textbf{ 0.3223**}           &0.3704 &0.2499 &0.2209 & \textbf{0.3575**}    &0.3555 &0.2289 & 0.2139& \textbf{0.3369**}\\
NDCG@20       &  0.3534            &0.3053               &  0.2813              &   \textbf{0.3319**}           &0.3784 &0.2579 &0.2243 &\textbf{0.3642**}    &0.3635 &0.2476 & 0.2141&\textbf{0.3352**}\\
NDCG@50       &   0.3946             &  0.3366            &   0.3219            &  \textbf{ 0.3719**}          &0.4169 &0.2918 &0.2484 &\textbf{0.4015**}    &0.4037&0.2808 &0.2288 &\textbf{0.3844**}\\   \bottomrule
\end{tabular}
\end{adjustbox}
\end{table*}

\subsubsection{\textbf{Compared Methods}}

We compare with the state-of-the-art machine unlearning methods on different recommendation models. The compared recommendation models include: 
\begin{itemize}
\item \textbf{BPR} \cite{rendle2009bpr}: This is a widely used recommendation model, which optimizes matrix factorization with the Bayesian Personalized Ranking objective function.
\item \textbf{WMF} \cite{hu2008collaborative,chen2020efficient2}: This is a non-sampling recommendation model, which treats all missing interactions as negative instances and weighting them uniformly. In our experiment, it is trained through the efficient non-sampling method proposed by \cite{chen2020efficient2}.
\item \textbf{LightGCN} \cite{he2020lightgcn}: This is a state-of-the-art graph neural network model which simplifies the design of GNN to make it more appropriate for recommendation.
\end{itemize}

The compared machine unlearning methods include:
\begin{itemize}
\item \textbf{Retrain}: The most straightforward machine unlearning method, which removes the revoked sample and retrains the model. It is added as the basic benchmark.
\item \textbf{SISA} \cite{bourtoule2021machine}: This is a state-of-the-art machine unlearning method, which randomly splits the training data into shards, then aggregates the results from all submodels through average for final prediction. 
\item \textbf{GraphEraser} \cite{chen2021graph}: This is a state-of-the-art machine unlearning method tailored to graph data, which uses node clustering methods for graph partition and a static weighted aggregation method for prediction.

\end{itemize}

\subsubsection{\textbf{Evaluation Methodology}}

For each dataset, we randomly select 80\% interactions to construct the training set and treat the remaining as the test set. 
From the training set, we randomly select 10\% interactions as the validation set to tune hyper-parameters. 
For each user, our evaluation protocol ranks all the items except the positive ones in the training set. To evaluate the recommendation performance, we apply two widely-used evaluation protocols: Recall@$N$ and NDCG@$N$. Recall@$N$ measures whether the ground truth is ranked among the top $N$ items, while NDCG@$N$ is a position-aware ranking metric, which assigns higher scores to higher positions.

\begin{table*}[]
\caption{Comparison results of running time (minute [m]). For RecEraser, only the corresponding submodel and the aggregation part need to be retrained when receiving an unlearning request of data.}
\label{table4}
\begin{tabular}{@{}ll r r r rrr rrr@{}}
\toprule
&              & \multicolumn{3}{c}{\textbf{Yelp2018}}  &\multicolumn{3}{c}{\textbf{Movielens-1m}} &\multicolumn{3}{c}{\textbf{Movielens-10m}}                                      \\\cmidrule(lr){3-5} \cmidrule(lr){6-8} \cmidrule(lr){9-11} 
\multicolumn{2}{c}{\multirow{2}{*}{}} &\textbf{BPR} &\textbf{WMF} & \textbf{LightGCN} &\textbf{BPR} &\textbf{WMF} & \textbf{LightGCN} &\textbf{BPR} &\textbf{WMF} & \textbf{LightGCN} \\ \midrule
\textbf{Retrain} &                  &  400m      &  75m      &  1090m$\quad$    &  330m      &  13m      & 545m$\quad$     &3250m          &  292m        &  25330m~~~     \\ \midrule
\multirow{3}{*}{\textbf{RecEraser}}   & \textbf{Shard Training}   &  39m      &   2.5m     &   52m$\quad$    &  12m      &  0.9m     &   17m$\quad$    &   240m       &  14m       &   610m~~~      \\
& \textbf{Aggregation Training}    &   13m     &  1.3m      & 34m$\quad$      &  10m      &   0.2m     &  14m$\quad$     &    80m      &  4m       &  330m~~~       \\
& \textbf{Total}    &   52m     &   3.8m     & 86m$\quad$      &  22m      & 1.1m       &  31m$\quad$     &    320m      &    16m     &  940m~~~       \\\bottomrule
\end{tabular}
\end{table*}
\subsubsection{\textbf{Hyper-parameter Settings}}

The hype-parameters are carefully tuned using a grid search to achieve optimal performances. 
In the case of model-specific hyper-parameters, we tune them in the ranges suggested by their original papers. After the tuning process, the batch size is set to 512, the learning rate is set to 0.05, the embedding size $d$ is set to 64, and the attention size $k$ is set to 32. The maximum number of epochs is set to 1000. The early stopping strategy is also adopted in our experiment, which terminates the training when Recall@10 on the validation set does not increase for 10 successive epochs.

\subsection{Recommendation Performance}

We first evaluate the recommendation performance of different unlearning methods. In this section, the proposed InBP method is used for RecEraser and the number of shards is set to 10 for all unlearning methods. 
To evaluate different Top-$N$ recommendation results, we set the length $N$ = 10, 20, and 50 in our experiments. The results are shown in Table \ref{table2}. From the table, the following observations can be made:

First, the retrain method generally can achieve the best performance since it uses the original dataset without the sample that need to be forgotten for training. However, as shown in the next section, its unlearning efficiency is low especially when the model is complex and the original dataset is large.

Second, our proposed RecEraser achieves significantly better performance ($p$ <0.01) than the state-of-the-art unlearning baselines on the three datasets. Specifically, on Yelp2018 dataset, RecEraser exhibits average improvements of 59\%, 41\%, and 46\% compared to SISA when using base models BPR, WMF, and LightGCN, respectively. Compared to GraphEraser, the above improvements are 11\%, 32\%, and 19\%, respectively. On other datasets, the results of RecEraser are also remarkable. The substantial improvement could be attributed to two reasons: (1) RecEraser partitions data through the proposed InBP method, which can enhance performance by preserving collaborative information; (2) RecEraser uses the proposed attention-based aggregation method to adaptively aggregate submodels, which could further improve the global model performance. 

Third, considering the performance on each dataset, we observe that the performances of unlearning methods depend on the sparsity of datasets. For example, the performances of RecEraser are more comparable with the retraining method on movielens-1m and movielens-10m since the two datasets are relatively dense than yelp2018. User preferences are more difficult to learn from sparse user–item interactions, thus the number of shards should be carefully set to balance the unlearning efficiency and performance. We make further discussions in Section \ref{sec5.4}.

\subsection{Unlearning Efficiency}
In this section, we conducted experiments to explore the unlearning efficiency of our proposed RecEraser. Following the settings of \cite{chen2021graph}, we randomly sample 100 user-item interactions from the training data, record the retraining time of the corresponding submodels and aggregation part, and calculate the average retraining time. The shard number is set to 10 in our experiment and all experiments in this section are run on the same machine (Intel Xeon 8-Core CPU of 2.4 GHz and single NVIDIA GeForce GTX TITAN X GPU). The results are shown in Table \ref{table4}. 
The experimental results show that compared to the retrain method, our proposed RecEraser can significantly improve the unlearning efficiency. For example, on movielens-10m dataset, our RecEraser only needs 320 minutes, 16 minutes, and 920 minutes to achieve the optimal performance for BPR, WMF, and LightGCN respectively, while the retrain method needs about 3250 minutes, 292 minutes, and 25330 minutes respectively. This acceleration is over 10 times, which is highly valuable in practice and is difficult to achieve with simple engineering efforts. For retrain method, training on a large data is very time-consuming. For our RecEraser, only the corresponding submodel and the aggregation part need to be retrained to forget the required data. Moreover, we can tolerate more shards for larger datasets to further improve the unlearning efficiency while preserving the performance.

\subsection{Ablation Study}

\subsubsection{\textbf{Effect of proposed data partition methods}}

\begin{figure}
\includegraphics[width=3.3in]{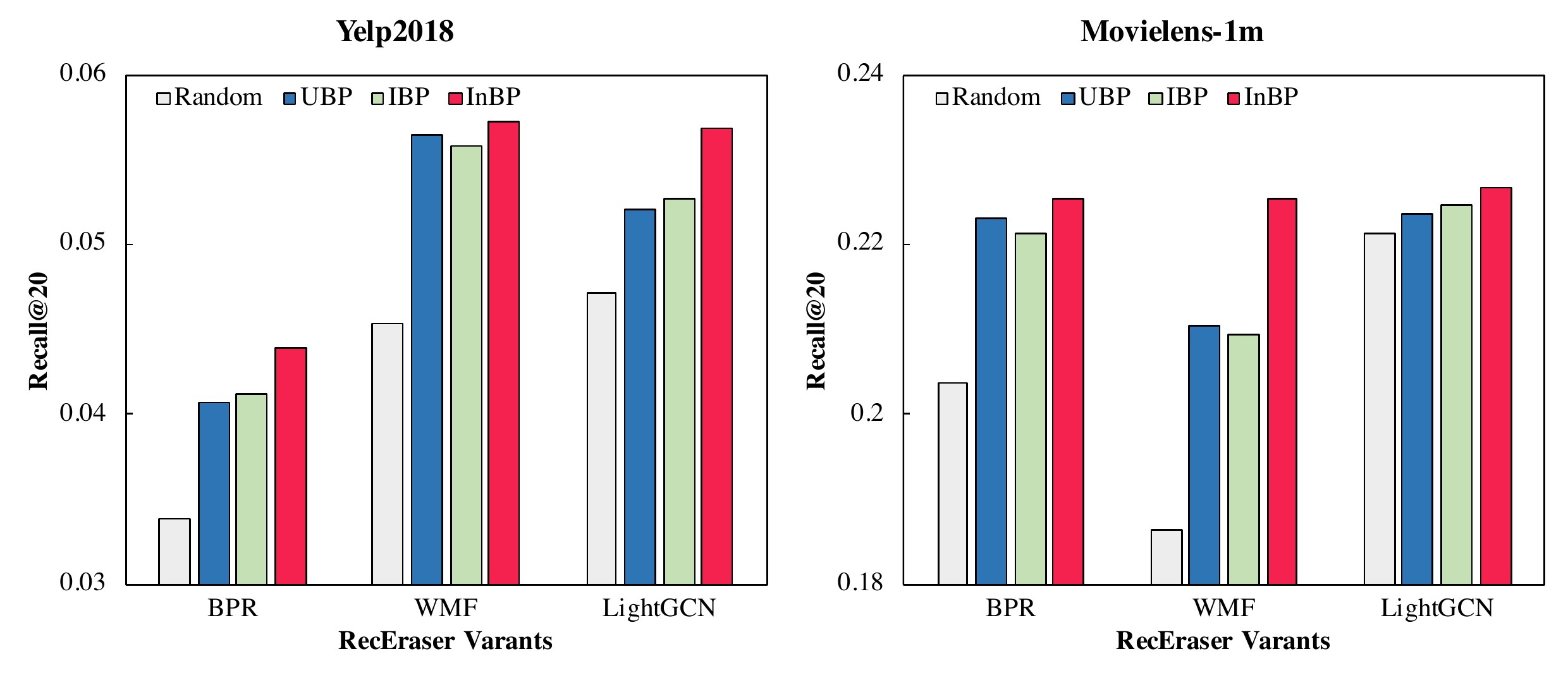}
\centering
\caption{Performance comparison of different data partition methods on Yelp2018 and Movielens-1m datasets.}
\label{figure2}
\end{figure}

In this paper, we propose three balanced data partition methods based on the similarities of users, items, and interactions. We first conduct ablation study to understand their effect. Figure \ref{figure2} shows the results w.r.t. Recall@20 of different methods on Yelp2018 and Movielens-1m datasets. 
The results of random partition are also shown as baselines. From Figure \ref{figure2}, two observations can be made: First, we can see that our proposed methods, UBP, IBP, and InBP, can achieve a much better recommendation performance compared to the Random method. Take the LightGCN trained on Yelp2018 dataset as an instance, the Recsall@20 for InBP is 0.0568, while the result is 0.0472 for Random. Second, from an overall view, InBP generally performs better than UBP and IBP, which indicates that it is more effective to partition the recommended dataset based on user-item interactions than based on users or items only.

\begin{figure}
\includegraphics[width=3.3in]{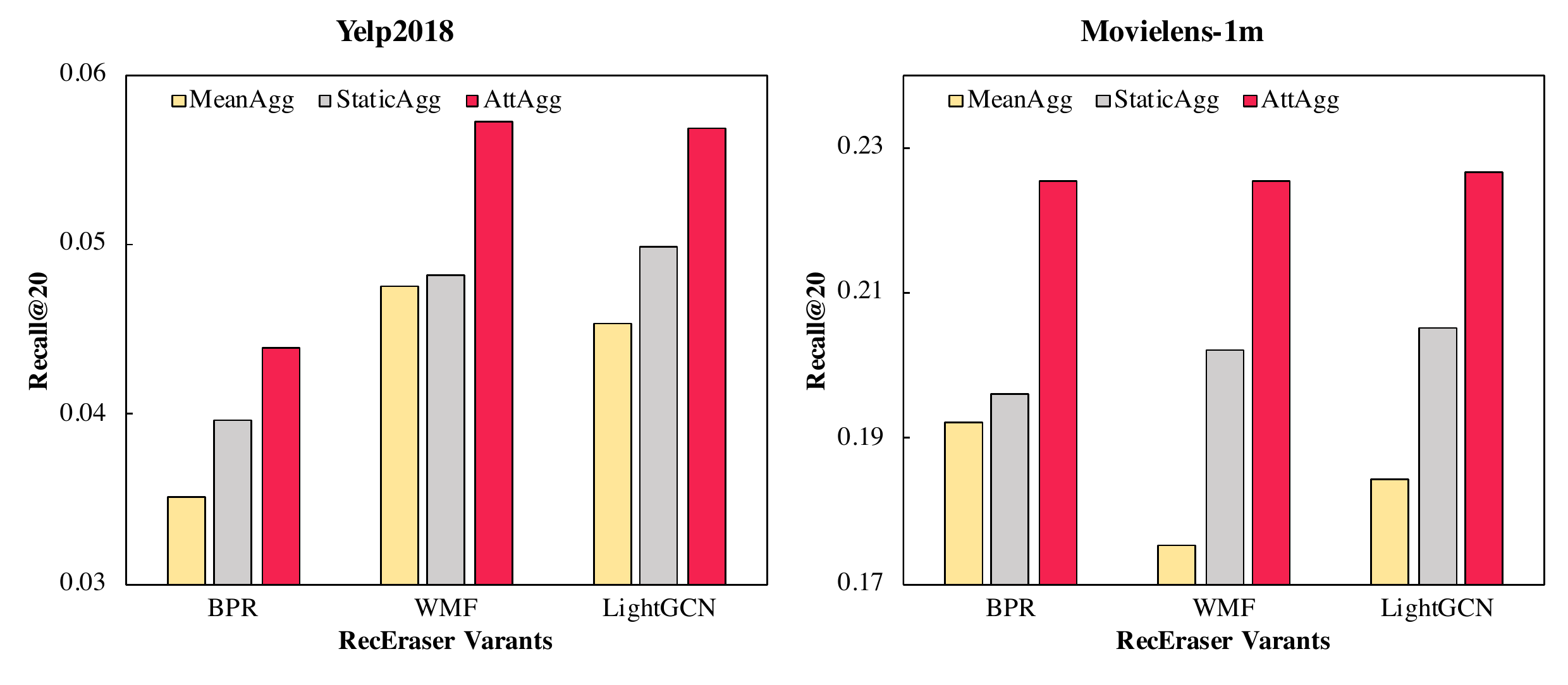}
\centering
\caption{Performance comparison of different data aggregation methods on Yelp2018 and Movielens-1m datasets.}
\label{figure3}
\end{figure}

\subsubsection{\textbf{Effect of proposed attention-based aggregation method}}

To illustrate the effectiveness of the proposed attention-based aggregation method (AttAgg), we compare with MeanAgg and StaticAgg on two datasets. MeanAgg has been utilized in SISA \cite{bourtoule2021machine}, which obtains the aggregated score by averaging the prediction of all submodels. 
StaticAgg has been utilized in GraphEraser \cite{chen2021graph}, which is a learning-based method to assign weights for different submodels. However, the weights can not changed for different user-item interactions. Figure \ref{figure3} shows the performance of the above methods. We can see that the proposed AttAgg method can effectively improve the recommendation performance compared to MeanAgg and StaticAgg. For example, on yelp2018 with WMF method, AttAgg achieves 0.0572 of Recall@20 score, while the corresponding result is 0.0475 for MeanAgg and 0.0482 for StaticAgg, respectively. 
This make sense since recommender systems usually face a variety of users and items, the learned features by each submodel should contribute differently for different predictions of users and items. Using AttAgg helps adaptively assign weights to different submodels when predicting different user-item pairs.

\subsection{Impact of the Shard Number}
\label{sec5.4}

\begin{figure}
\includegraphics[width=3.3in]{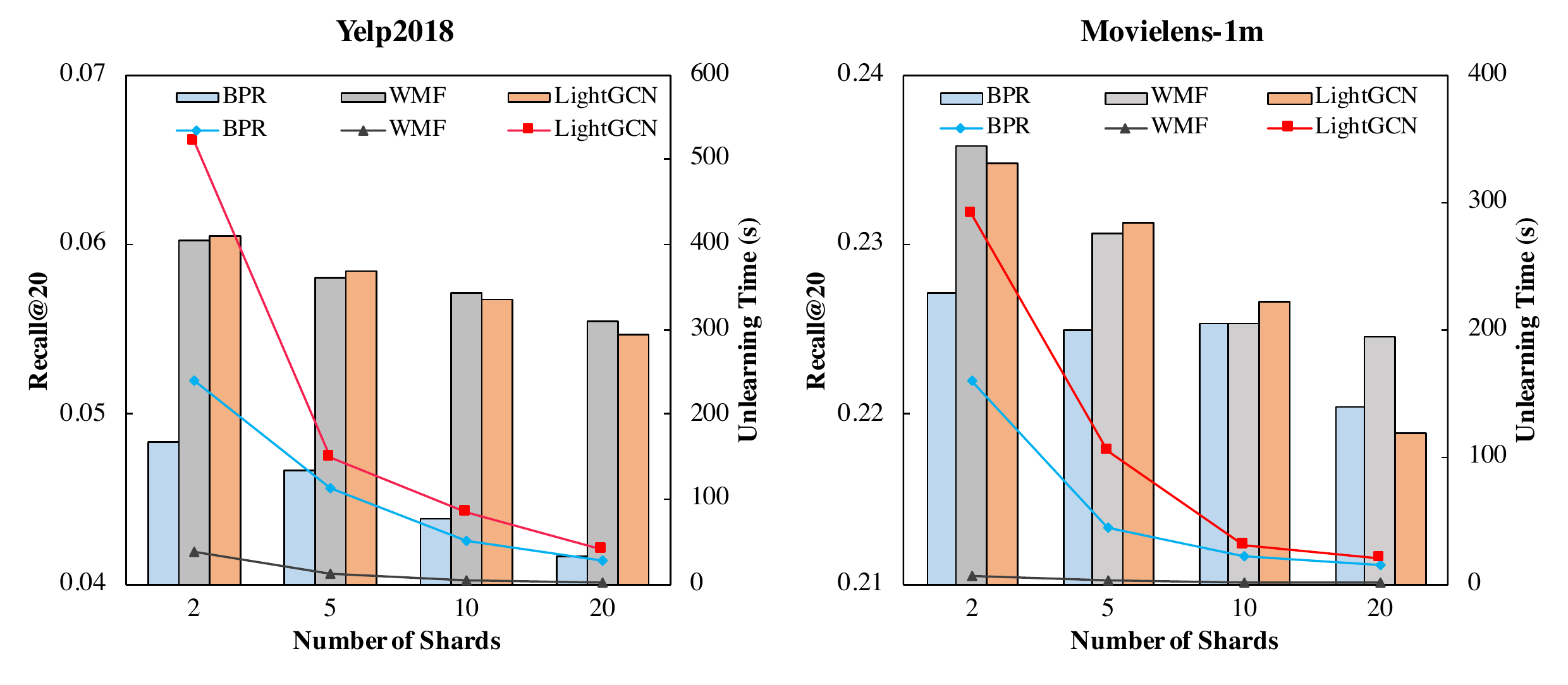}
\centering
\caption{Impact of the shard number on the unlearning efficiency and performance on Yelp2018 and Movielens-1m datasets. The bar chart shows the recommendation performance and the line chart shows the unlearning time cost. }
\label{figure4}
\end{figure}

We conduct the experiments on Yelp2018 and Movielens-1m datasets to investigate the impact of shard number.
The experimental results in Figure \ref{figure4} show that the average unlearning time decreases when the number of shards increases for all recommendation models. This make sense as a larger number of shards means smaller shard size, which will improve the unlearning efficiency. Meanwhile, the recommendation performance of all the three recommendation models slightly decrease. 
This is because recommendation models require collaborative information for model learning, which generally decreases as the number of shards increases. In practice, The number of shards should be carefully selected to balance the unlearning efficiency and recommendation performance.

\section{Conclusion and Future Work}

In this paper, we propose a novel RecEraser framework, which is, to the best of our knowledge, the first machine unlearning method tailored to recommendation tasks.
To permit efficient unlearning while keeping the collaborative information of the data, we first design three data partition strategies. We then further propose an adaptive aggregation method to improve the global model utility. Extensive experiments have been made on three real-world datasets and three representative recommendation models. The proposed RecEraser can not only achieve efficient unlearning, but also outperform the state-of-the-art unlearning methods in terms of model utility. We believe the insights of RecEraser are inspirational to future developments of recommendation unlearning methods. With the prevalence of data protection requirements in real applications, erasable models are becoming increasingly important in recommendation. Our RecEraser also has the potential to benefit many other tasks where the collaborative information should be considered.

In this work, we focus on the setting where the unlearning requests arrive sequentially to study the feasibility of unlearning in recommender systems.
Alternatively, the system can also aggregate several sequential unlearning requests into a batch and then unlearn this batch at once.
At present, how to efficiently unlearning using SISA framework in batch setting is still an open problem~\cite{bourtoule2021machine}.
In the future, we seek to improve the unlearning efficiency for recommendation in batch setting.
In addition, we will also try to incorporate the content features of users and items, and exploring our method in other related tasks like social network embedding. 

\begin{acks}
This work is supported by the Natural Science Foundation of China (Grant No. U21B2026) and Tsinghua University Guoqiang Research Institute. 
\end{acks}

\balance
\bibliographystyle{ACM-Reference-Format}
\bibliography{unlearning}

\end{document}